\documentclass{iopart}
\usepackage{graphicx}
\usepackage{iopams}
\begin{document}

\title[A Source to the Spherical Antennas]
{Detectability of f-mode unstable Neutron Stars by the Schenberg
Spherical Antenna}

\author[de Araujo et al]{Jos\'{e} Carlos N de Araujo$^{1}$,
Oswaldo D Miranda$^{1,2}$ and Odylio~D~Aguiar$^{1}$}

\address{$^{1}$Instituto Nacional de Pesquisas Espaciais -
Divis\~ao de Astrof\'\i sica \\ Av. dos Astronautas 1758, S\~ao
Jos\'e dos Campos, 12227-010 SP, Brazil}

\address{$^{2}$Instituto Tecnol\'{o}gico de Aeron\'{a}utica -
Departamento de F\'\i sica \\ Pra\c{c}a Marechal Eduardo Gomes 50,
S\~ao Jos\'e dos Campos, 12228-900 SP, Brazil}

\ead{\mailto{jcarlos@das.inpe.br, oswaldo@das.inpe.br,
odylio@das.inpe.br}}

\begin{abstract}
The Brazilian spherical antenna (Schenberg) is planned to detect
high frequency gravitational waves (GWs) ranging from 3.0 kHz to
3.4 kHz. There is a host of astrophysical sources capable of being
detected by the Brazilian antenna, namely: core collapse in
supernova events; (proto)neutron stars undergoing hydrodynamical
instability; f-mode unstable neutron stars, caused by quakes and
oscillations; excitation of the first quadrupole normal mode of
4--9 solar mass black holes; coalescence of neutron stars and/or
black holes; exotic sources such as bosonic or strange matter
stars rotating at 1.6 kHz; and inspiralling of mini black hole
binaries. We here address our study in particular to the neutron
stars, which could well become f-mode unstable producing therefore
GWs. We estimate, for this particular source of GWs, the event
rates that in principle can be detected by Schenberg and by the
Dutch Mini-Grail antenna.
\end{abstract}

%\submitto{\CQG}


% 04.30.Db - wave generation and sources
% 04.80.Nn - gravitational wave detectors and experiments
% 95.55.Ym - gravitational radiation detectors; ...
% 97.60.Gb - Pulsars
% 97.60.Jd - Neutron stars
% 97.60.Lf - Black holes

%\maketitle

\section{Introduction}

The Brazilian antenna is made of Cu-Al (94$\%$-6$\%$), it has a
diameter of 65 cm and covers the 3.0-3.4 kHz bandwidth using a
two-mode parametric transducer (see, Aguiar \etal 2002, 2004 for
details).

The initial target is a sensitivity of $\tilde{h}\sim 2\times
10^{-21}\, {\rm Hz^{-1/2}}$ in a 50 Hz bandwidth, which we believe
is reachable at 4.2 K with conservative parameters. An advanced
sensitivity will be pursued later, cooling the antenna to 15 mK
and improving all parameters (see Frajuca \etal 2004).

The Brazilian antenna will operate in conjunction with the Dutch
Mini-GRAIL antenna, and the laser interferometer detectors, which
can also cover such high frequencies with similar sensitivities.

It is worth mentioning that the event rates for the two spherical
antennas will be the same, as long as they have the same
sensitivity. So, throughout the paper, when we refer to the
sources and the event rates to the Schenberg antenna, the reader
has to bear in mind that we are also refereing to the the Dutch
Mini-Grail antenna.

In a previous work by a M.Sc. student of our group (Castro 2003),
a preliminary study of the most probable sources of gravitational
waves (GWs) was conducted that could be detected by Schenberg,
namely: core collapse in supernova events; (proto)neutron stars
undergoing hydrodynamical instability; f-mode unstable neutron
stars, caused by quakes and oscillations; excitation of the first
quadrupole normal mode of $4-9 M_{\odot}$ black holes; coalescence
of neutron stars and/or black holes; exotic sources such as
bosonic or strange matter stars rotating at 1.6 kHz; and
inspiralling of mini black hole binaries.

Our aim here is to present the results of this study and to
revisit the $f-$mode unstable neutron star source of GWs. This
specific pulsation mode is one of the most important channel for
the emission of GWs (Kokkotas and Andersson 2001) by neutron
stars. In particular, we are using recent results for the radial
distribution of pulsars in the Galaxy (Yusifov and
K\"{u}\c{c}\"{u}k 2004) in order to determine the event rate
detectable by Schenberg for a given efficiency of generation of
GWs.

The paper is organized as follows. In section 2 we briefly
consider the sources detectable by the Schenberg antenna, in
section 3 we revisit the f-mode neutron star GW detectability by
the Schenberg antenna, and finally in section 4 we present our
conclusions.

\section{The Sources to Schenberg}

First of all, it is worth mentioning that in the present estimates
of event rates we are assuming that the Schenberg's sensitivity to
burst sources is of $h \sim 10^{-20}$, which seems reasonable from
our projected $\tilde{h}$ and bandwidth.

It is important to bear in mind that such a sensitivity is not the
quantum limit one, which could be a factor around 5 better. Also,
it is worth remembering that using the ``squeezing technique" the
quantum limit could in principle be overtaken. All this would
imply that Schenberg, which will be using parametric transducers,
could in principle present event rates significantly greater than
those presented here; in some cases, the rates could well be
higher by a factor of up to $\sim 10^{2}$.

It is worth stressing that the Brazilian detector will be
sensitive to sources of the local group of galaxies ($r\sim
1.5{\rm Mpc}$). Although, just the Galaxy, ${\rm M}31$ and ${\rm
M}33$ can give a significant contribution to the event rates,
because these three galaxies account for more than $90\%$ of the
mass of the local group.

Our estimates show that, except for the mini black hole binaries
and the f-mode unstable neutron stars, the other putative sources
to Schenberg present event rates of at most one event every $\sim$
10 years, at a signal-to-noise ratio (SNR) equal to unity. Thus,
the prospect for the detection of these sources is not very
promising. Because of this we do not enter into details of such
estimates.

For the mini black hole binaries we refer the reader to the paper
by de Araujo \etal (2004) for details. They show that the event
rate in this case may be of one event every 5 years, at SNR equal
to unity.

Our main aim in the present paper is to pay attention to the
f-mode unstable neutron star, which can in principle be an
important source to the Schenberg antenna. In our previous study
we show that one such an event every year would be detectable by
Schenberg at a SNR equal to unity.

In the next section we revisit the f-mode unstable neutron star
study concerning its detectability in particular by the Schenberg
antenna.

\section{GWs from f-Mode Unstable Neutron Stars}

Before studying the f-mode unstable neutron star as a source of
GWs to the Brazilian antenna, which is of major interest here, it
is worth considering its relevance as compared to the other
pulsation modes such stars could have in what concern the
generation of GWs.

Relativistic stars are known to have a host of pulsation modes.
Only a few of them, however, is of relevance for GW detection.
From the GW point of view the most important modes are the
fundamental (f) mode of fluid oscillation, the first few pressure
(p) modes and the first GW (w) mode (Kokkotas and Schutz 1992).
Among these three modes the pulsation energy is mostly stored in
the f-mode in which the fluid parameters undergo the largest
changes. It is worth mentioning that the r-mode can also be, under
certain circumstances, a very important source of GWs (Andersson
\etal 2001).

An important question is how the modes are excited in the neutron
stars, which are of our concern here. There are many scenarios
that could lead to significant asymmetries. A supernova explosions
are expected to form a wildly pulsating neutron star that emits
GWs. A pessimistic estimate for the energy radiated as GWs
indicates a total release equivalent to $< 10^{-6}M_{\odot}$. An
optimistic estimates, where the neutron star is formed, for
example, from strongly non-spherical collapse, suggest a release
equivalent to $10^{-2}M_{\odot}$.

Another possible excitation mechanism for neutron star pulsation
is a starquake, which can be associated with a pulsar glitch. The
energy released in this process may be of the order of the maximum
mechanical energy stored in the crust of the neutron stars, which
is estimated to be of $10^{-9} - 10^{-7}M_{\odot}$ (Blaes \etal
1989; Mock and Joss 1998).

During the coalescence of two neutron stars several oscillation
modes could in principle be generated. Stellar oscillations  being
excited by the tidal fields of the two stars, for example.

The neutron star may undergo a phase transition leading to a
mini-collapse, which could lead to a sudden contraction during
which part of the gravitational binding energy of the star would
be released, and, as a result, it could occur that part of this
energy would be channelled into pulsations of the remnant.
Similarly, the transformation of a neutron star into a strange
star is likely to induce pulsations.

In our previous study we have found that Schenberg could in
principle detect at least one such a source per year at SNR equal
to unity. The basic assumptions in this study are the following.

Firstly, we have taken into account in our estimate only the known
Pulsars. Secondly, we have assumed that the energy release in GWs
is of the order of  $10^{-6} M_{\odot}$ (see, e.g., Andersson and
Kokkotas 1996). Thirdly, we have associated the f-mode excitation
with the same mechanism responsible to the glitch phenomenon,
which is related to some neutron star internal structure
rearrangement (see, e.g., Horvath 2004).

Last but not least, we have assumed that the f-mode instability
may produce GWs in the frequency band of $3.0-3.4$ kHz, that of
Schenberg's. We refer the reader to the paper by Kokkotas and
Andersson (2001), in particular its figure 2, where it is shown
clearly that for a family of equations of state (EOSs) GWs of
$\sim 3$ kHz may be produced by f-mode unstable neutron stars.

It is worth recalling that GWs produced in the f-mode excitation
depends on the EOS for the neutron star matter that, as is well
known, is not completely established.

Before considering the improvements we intend to take into account
in revisiting this study, it is worth recalling that the
characteristic amplitude of GWs related to the f-mode instability
is given by
\begin{equation} \label{h}
h \simeq 2.2\times 10^{-21}\left(\frac{\varepsilon_{GW}}{10^{-6}
}\right)^{1/2}\left(\frac{2\,
kHz}{f_{GW}}\right)^{1/2}\left(\frac{50\, kpc}{r} \right),
\end{equation}
\noindent (see, e.g., Andersson and Kokkotas 1998) where
$\varepsilon_{GW}$ is the efficiency of generation of GWs,
$f_{GW}$ is the GW frequency, and $r$ is the distance to the
source.

It is worth mentioning that the f-mode is a burst source of GWs
with a duration of tenths of a second (see, e.g., Andersson and
Kokkotas 1998). This signal being concentrated in a bandwidth
which is completely within the Schenberg's bandwidth. As a result
in the calculation of the SNR it is a good approximation to
compare directly the detector pulse sensitivity with the predicted
wave amplitude of the f-mode GW.

The Schenberg's sensitivity for burst sources can be of the order
of $10^{-20}$. This implies that, for $\varepsilon_{GW} \sim
10^{-6}$ and $f_{GW}=$3 kHz, Schenberg can in principle detect
f-mode unstable neutron star sources at distances of up to $r \sim
10$ kpc at $SNR\sim 1$.

Certainly, the number of neutron stars within the volume {\it
seen} by Schenberg could be in principle enormous. Unless the
efficiency of generation of GWs through f-mode instability is $
\ll 10^{-6}$ or such a mode is not excited at all, Schenberg could
in principle detect f-mode unstable neutron stars with a
considerable event rate.

In this study the main ingredient we have taken into account is
the distribution function of pulsars in the Galaxy. One could
consider, however, that it would be desirable to take into
account, instead, a distribution function for the neutron stars in
the Galaxy, because the pulsar population is a tiny part (say
0.1-0.01\%; later on we explain how these figures are obtained) of
the neutron star population. But, one has to bear in mind that
most of the observed neutron stars is in fact seen in the form of
pulsars.

It is worth mentioning that if the f-mode instability occurs in
any neutron star, being it a pulsar or not, the event rate seen by
any GW detector sensitive to its frequency could be strongly
enhanced.

The distribution function for the pulsars in the Galaxy has been
studied in many papers (see, e.g., Narayan 1987, Paczynski 1990,
Hartman \etal 1997; Lyne \etal 1998; Schwarz and Seidel 2002,
among others).

We here adopt the distribution function recently obtained by
Yusifov and K\"{u}\c{c}\"{u}k (2004), namely,
\begin{equation}
\rho(R)=A\biggl({R \over R_\odot} \biggr)^a
\exp{\biggl[-b\biggl({R-R_\odot \over R_\odot }\biggr) \biggr] }
\label{Gam14} ,
\end{equation}
\noindent where $\rho(R)$ is the surface density of pulsars, $R$
is Galactocentric distance, $R_\odot=8.5$ kpc is the
Sun$-$Galactic center (GC) distance. Note that, the equation
(\ref{Gam14}) implies that $\rho(0)=0$, which is inconsistent with
observations. To avoid such a problem the authors include an
additional parameter $R_1$ and used a shifted Gamma function,
replacing $R$ and $R_\odot$  in equation~(\ref{Gam14}) by
$X=R+R_1$ and $X_\odot =R_\odot +R_1$, respectively. The best fit,
using the LMS method gives: $A=37.6\pm1.9 {\rm kpc}^{-2}, \>
a=1.64\pm0.11$, $b=4.01\pm0.24$ and $R_1=0.55\pm0.10$ kpc. We
refer the reader to the paper by Yusifov and K\"{u}\c{c}\"{u}k
(2004) for further details.

In figure 1 we present the number of pulsars as a function of the
distance from the sun, which has been obtained through integration
of equation~(\ref{Gam14}). Also present is the number of pulsars
corrected by the beaming factor, which multiples that number by a
factor of approximately 10 (see, e.g., Tauris and Manchester
1998).

The number of pulsars that could be seen by Schenberg amount $\sim
10^{5}$ (taking into account the beaming correction). In the whole
Galaxy the number of pulsars with luminosity greater than 0.1~mJy
${\rm kpc^{2}}$ at 1400MHz is predicted to be of $\sim 2.4\times
10^{5}$, taking into account the beaming factor. Again, we refer
the reader to the paper by Yusifov and K\"{u}\c{c}\"{u}k (2004)
for further details.

\begin{figure}
\begin{center}
\includegraphics[width=6cm]{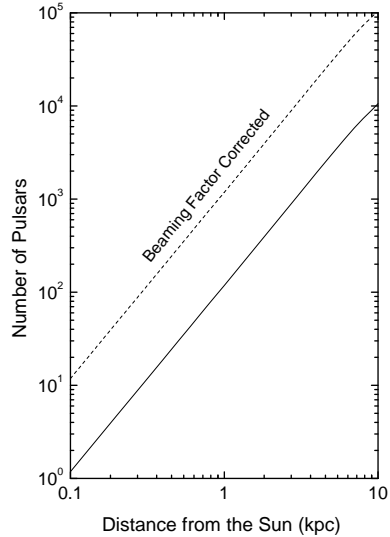}
\caption{Number of pulsars as a function of the distance from the
sun without (solid line) and with (dashed line) the beaming factor
correction.}
\end{center}
\end{figure}

Now, from the available catalogs at that time, Castro (2003)
obtained that only $3\%$ of the known pulsars present glitch
phenomenon. The number of glitches in 25 years amounting to 45.
From Castro we obtain that the number of glitches per year per
pulsars amount to $2.6\times 10^{-3}$.

Since we are considering the fact that the f-mode is capable of
being excited during the glitch phenomenon we have:
\begin{equation}
2.6\times 10^{-3} events/yr/pulsar.
\end{equation}

Finally, to obtain the number of events per year, detectable by
Schenberg, we use the results presented in figure 1.

Since the efficiency of generation of GW through f-mode channel is
not known, we present in figure 2 the event rate detectable by
Schenberg as a function of $\varepsilon_{GW}$. Note that for
$\varepsilon_{GW} > 10^{-8}$ ($10^{-7}$) we predict that one
f-mode source could in principle be detected at $SNR=1$ ($SNR=3$)
every year. It is worth noting that the number of pulsars used in
the calculation of figure 2 takes into account the beaming
correction.

\begin{figure}
\begin{center}
\includegraphics[width=6cm]{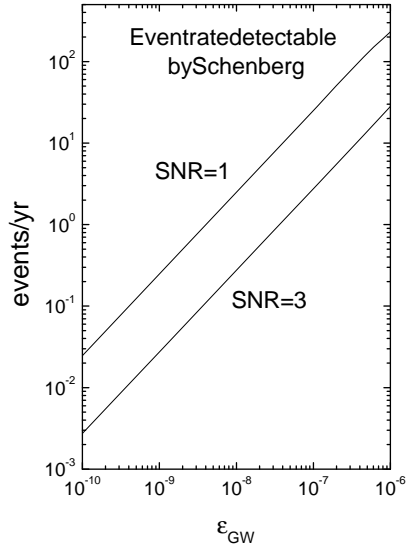}
\caption{Number of events per year, i.e., number of f-mode
unstable pulsars per year, as a function of $\varepsilon_{GW}$,
the efficiency of generation of GWs, detectable by Schenberg at
$SNR=1$ and $3$, assuming that the Schenberg's sensitivity to
burst sources is of $h \sim 10^{-20}$.}
\end{center}
\end{figure}

Note that the event rate is critically dependent on the value of
$\varepsilon_{GW}$. In figure 2 one sees that if this parameter is
too small the event rate is very small too.

However, the prediction appearing in figure 2 takes into account
only the neutron stars in the form of pulsars with luminosity
greater than 0.1~mJy ${\rm kpc^{2}}$ at 1400MHz. The number of
neutron stars in the Galaxy, pulsars or not, could well be a
factor of a thousand, or even tens of a thousand, greater.

Paczynski (1990), for example, estimates that there may exist
$\sim 10^{9}$ neutron stars in the Galaxy. Other authors find
similar figures, namely: Nelson \etal (1995) and Walter (2001)
argue that there may exit $\sim 10^{8} - 10^{9}$ neutron stars in
the Galaxy; whereas, Timmes \etal (1996), using models for massive
stellar evolution coupled to a model for Galactic chemical
evolution, obtained $\sim 10^{9}$ neutron stars in the Galaxy.

If the non pulsar neutron stars could also be f-mode unstable,
this means that the event rate detectable by Schenberg could be
greatly enhanced. In particular, if the fraction of neutron stars
and the fraction of pulsars, which are f-mode unstable, are
similar this means that the event rate that could be detected by
Schenberg would be a thousands, or even a few thousands, greater.

If this is the case, even though $\varepsilon_{GW} \sim 10^{-10}$,
Schenberg could detect $\sim 10-100$ events every year at $SNR=3$.

\section{Final Remarks}

Particular attention has been given here to the f-mode sources,
because it can in principle be one of the most important
candidates to be detected by the Schenberg antenna, with an event
rate that could amount to several sources every year.

Since the interferometers are also sensitive to the GWs generated
by f-mode neutron stars, it would be of interest to search for
this sources with such detectors. Due to the fact that the
sensitivity of the interferometers at 3 kHz (see, e.g., Shoemaker
2005) could well be similar to that of the Schenberg antenna, the
event rate of both detectors could be similar.

Also, it is worth mentioning that the interferometers could probe
a wider range of EOSs, as compared to the Schenberg antenna, since
they are sensitive to broader GW frequency band.

Finally, it is worth mentioning that Kokkotas \etal (2001) show
that detecting the f-mode, the EOS, the mass and the radius of the
neutron stars will be strongly constrained. The reader should
appreciate the reading of this paper by Kokkotas \etal, who show
in detail how these above mentioned astrophysical information is
obtained from the GW data.

\ack{JCNA would like to thank FAPESP (grant 2004/04366-0) and CNPq
(grant 304666/2002-5) for financial support; ODM would like to
thank FAPESP (02/01528-4 and 02/07310-0) for financial support;
ODA would like to thank FAPESP (1998/13468-9)and CNPq
(306467/03-8) for financial support.}

\References
\item[] Aguiar O D \etal 2002 \CQG {\bf 19} 1949

\item[] Aguiar O D \etal 2004 \CQG {\bf 21} S457

\item[] Andersson N and Kokkotas K D 1996 \PRL {\bf 77} 4134

\item[] Andersson N and Kokkotas K D 1998 {\it Mon. Not. R.
Astron. Soc.} {\bf 299} 1059

\item[] Blaes O, Blandford R, Goldreich P, Madau P 1989 {\it
Astrophys. J} {\bf 343} 839

\item[] Castro C S 2003 {\it Master thesis} (S.J. Campos:
INPE-10118-TDI/896)

\item[] de Araujo J C N, Miranda O D, Castro C S, Paleo B W and
Aguiar O D 2004 \CQG {\bf 21} S521

\item[] Frajuca C, Ribeiro K L, Andrade L A, Aguiar O D,
Magalh\~{a}es N S and de Melo Marinho Junior R 2004 \CQG {\bf 20}
S1107

\item[] Hartman J W \etal 1997 {\it Astron. Astr.} {\bf 322} 477

\item[] Horvath J E 2004 {\it Int. J. Mod. Phys.} {\bf D 13} 1327

\item[] Kokkotas K D and Andersson N 2001 {\it Preprint
gr-qc/0109054}

\item[] Kokkotas K D, Apostolatos A T and Andersson N 2001 {\it
Mon. Not. R. Astron. Soc.} {\bf 320} 307

\item[] Kokkotas K D and Schutz B F 1992 {\it Mon. Not. R. Astron.
Soc.} {\bf 255} 119

\item[] Lyne A G \etal 1998 {\it Mon. Not. R. Astron. Soc.} {\bf
295} 743

\item[] Mock P C and Joss P C 1998 {\it Astrophys. J.} {\bf 500}
374

\item[] Narayan R 1987 {\it Astrophys. J.} {\bf 319} 162

\item[] Nelson R W, Wang J C L, Salpeter E E, Wasserman I 1995
{\it Astrophys. J.} {\bf 438} L99

\item[] Paczynski B 1990 {\it Astrophys. J.} {\bf 348} 485

\item[] Schwarz D J and Seidel D 2002 {\it Astron. Astr.} {\bf
388} 483

\item[] Shoemaker D \etal 2005 \CQG {\bf 22} in press

\item[] Timmes F X, Woosley S E and Weaver T A 1996 {\it
Astrophys. J.} {\bf 457} 83

\item[] Tauris T M and Manchester R N 1998 {\it Mon. Not. R.
Astron. Soc.} {\bf 298} 625

\item[] Walter, F M  2001 {\it Astrophys. J.} {\bf 549} 433

\item[] Yusifov I and K\"{u}\c{c}\"{u}k I 2004  {\it Astro. Astr.}
{\bf 422} 545

\endrefs

\end{document}